\documentclass[times,a4]{oupau}
\usepackage{natbib}
\newcommand{\eps}{\epsilon}

\newcommand{\RR}{{\mathbb R}}
\newcommand{\IR}{{\mathbb R}}

\newcommand{\CIc}{{\mathcal C}^\infty_{\rm{c}} }

\newcommand{\cO}{{\mathcal O}}
\newcommand{\cP}{{\mathcal P}}

\newcommand{\supp}{\operatorname{supp}}

\newcommand{\rest}{|}

\renewcommand{\Re}{\mathop{\rm Re}\nolimits}
\renewcommand{\Im}{\mathop{\rm Im}\nolimits}

\newcommand{\Id}{\operatorname{Id}}

\newcommand{\be}{\begin{equation}}
\newcommand{\ee}{\end{equation}}

\newcommand{\defeq}{\stackrel{\rm{def}}{=}}
\def\hto0{\xrightarrow{h\to 0}}

\theoremstyle{plain}

\newtheorem{lem}{Lemma}[section]

\theoremstyle{definition}

\numberwithin{equation}{section}

\newcommand{\bequ}{\begin{equation}}

\newcommand{\norm}[1]{\Vert#1\Vert}

\begin{document}

\runningheads{S.~Nonnenmacher and M.~Zworski}{Semiclassical resolvent estimates}

\title{Semiclassical resolvent estimates in chaotic scattering}
\author
{St\'ephane Nonnenmacher\affil{a}\corrauth and Maciej Zworski \affil{b}}
\address{\affilnum{a}Institut de Physique Th\'eorique, 
CEA/DSM/PhT, Unit\'e de recherche associ\'ee au CNRS,
91191 Gif-sur-Yvette, France\\
\affilnum{b}Mathematics Department, University of California,
Evans Hall, Berkeley, CA 94720, USA}
%\email{zworski@math.berkeley.edu}
\corraddr{Institut de Physique Th\'eorique, 
CEA-Saclay,
91191 Gif-sur-Yvette, France. Email:snonnenmacher@cea.fr}

\begin{abstract}
We prove resolvent estimates for semiclassical operators such as $-h^2 \Delta+V(x)$ in scattering
situations. Provided the set of trapped classical trajectories supports
a chaotic flow and is sufficiently filamentary, the analytic continuation of the 
resolvent is bounded by $h^{-M}$ in a strip
whose width is determined by a certain topological pressure associated with the classical flow. 
This polynomial estimate
has applications to local smoothing in Schr\"odinger propagation and to energy decay of
solutions to wave equations.
\end{abstract}
\keywords{Quantum scattering, chaotic trapped set, semiclassical resolvent estimates}

\received{}

\maketitle

%%%%%%%%%%%%%%%%%%%%%%%%%%%%%%%%%%%%%%%%%%%%%%%%%%%%%%%%%%%%%%%%%%%%%%%%%%%%%%%
%%%%%%%%%%%%%%%%%%%%%%%%%%%%%%%%%%%%%%%%%%%%%%%%%%%%%%%%%%%%%%%%%%%%%%%%%%%%%%%

\section{Statement of Results}
\label{in}

In this short note we prove a resolvent estimate in 
the pole free strip for operators whose 
classical Hamiltonian flows are hyperbolic on the sets of trapped trajectories (trapped sets),
and the latter are assumed to be 
sufficiently filamentary -- see \eqref{eq:pre} for the precise condition. 
The proof is based on the arguments of \cite{NZ3}
and we refer to \S 3 of that paper for the 
preliminary material and assumptions on the operator.

The polynomial estimate on the resolvent in the pole free strip below the real axis
\eqref{eq:reses} provides a direct proof of the estimate
on the real axis \eqref{eq:blog}, and that estimate is only logarithmically 
weaker than the similar bound in the non-trapping case (that is, the case where all 
classical trajectories escape to infinity). 
Through an argument going back to Kato, and more recently
to Burq,  
that estimate is crucial for obtaining local smoothing and Strichartz
estimates for the Schr\"odinger equation. These in turn are
important in the investigation of nonlinear waves in non-homogeneous 
trapping media. Also, as has been known since the work of 
Lax-Phillips, the estimate in the complex domain is useful for
obtaining exponential decay of solutions to wave equations (see
the paragraph following \eqref{eq:blog} for some references to recent
literature).

An example of an operator to which our methods apply is 
given by the semiclassical Schr\"odinger operator
\begin{equation}
\label{eq:Pu}
 P u(x) =  P ( h ) u(x) = - h^2 \frac{1}{\sqrt {\bar g } }
\sum_{i,j=1}^n \partial_{x_j} \left( \sqrt {\bar g} g^{ij} \partial
_{x_i} u(x) \right) + V ( x ) \,, \ \  x\in \RR^n \,, 
\end{equation}
 $ G ( x ) \defeq (g^{ij} ( x ))_{i,j} $ 
is a symmetric positive definite matrix representing a (possibly nontrivial) metric
on $\IR^n$, $ \bar g \defeq 1/ \det G ( x) $, and $V(x)$ is a potential function.
We assume that the geometry and the potential are ``trivial'' outside a bounded region:
\[ g^{ij} ( x ) = \delta_{ij} \,, \ \  V( x ) = - 1 \,, \ \ \text{when}\ |x| > R \,.\]
This operator is hence associated with a short-range scattering situation.
We refer to \cite[\S 3.2]{NZ3} for the complete set of assumptions
which allow long range perturbations, at the expense of some 
analyticity assumptions% standard for the definition of resonances. -- see \cite{Sj} and references given there. 
We note that for $ V \equiv - 1 $, $P(h)u-0$ is the Helmholtz
equation for a Laplace-Beltrami operator, with $ h = 1/\lambda $, 
playing the r\^ole of wavelength. 
%The Euclidean space can be 
%without any changes replaced by a smooth manifold coinciding with 
%$ (\RR^n \setminus B( 0 , R ) ) \sqcup \cdots \sqcup 
%(\RR^n \setminus B ( 0 , R ) )  $ outside of a compact set.

Such operators have a purely continuous spectrum near the origin, and their
truncated resolvent $\chi(P(h)-z)^{-1}\chi$ ($\chi\in C^\infty_c(\IR^n)$) 
can be meromorphically continued from $\Im z>0$ to $\Im z<0$, with poles of finite
multiplicity called {\it resonances}. In the semiclassical limit $h\ll 1$, the distribution
of resonances depends on the properties of the classical flow
generated by the Hamiltonian 
$$
p ( x , \xi ) = \sum_{i,j=1}^n g^{ij} ( x ) \xi_i \xi_j + V ( x ), 
$$
that is the flow $(x,\xi)\mapsto \exp  t H_p ( x , \xi )$ associated with the 
Hamiltonian vector field 
$$
H_p ( x, \xi ) \defeq \sum_{k=1}^n  \partial_{ \xi_k } p  \, \partial_{ x_k } - 
\partial _{x_k } p \, \partial_{ \xi_k }\,.
$$
(when $ V \equiv -1 $ the Hamiltonian flow corresponds to the geodesic flow on 
$ S^* \RR^n $.) 
More precisely, the properties of the resolvent $\chi(P(h)-z)^{-1}\chi$ near $z=0$ are influenced by
the nature of flow on the energy shell $\{p(x,\xi)=0\}$. A lot of attention has been 
given to {\it nontrapping flows}, that is flows for which the {\it trapped set}
\begin{equation}\label{e:trapped}
K \defeq \{ ( x, \xi ) \; : \; p ( x , \xi ) = 0 \,, \ \ 
\exp  t H_p ( x , \xi ) \not \rightarrow \infty \,, \ t \rightarrow 
\pm \infty \}
\end{equation}
is empty. In 
that case, for $\delta>0$ small enough and any $C>0$, 
the resolvent is pole free in a strip $[-\delta,\delta]-i [0,Ch]$, and 
satisfies the bound \cite{Mar02,NaStZw03}
$$
\|\chi(P(h)-z)^{-1}\chi\|_{L^2\to L^2}=\cO(h^{-1}),\quad  z\in [-\delta,\delta]-i [0,Ch]\,.
$$
On the opposite, there exist cases of ``strong trapping'' for which the trapped set
has a positive volume; resonances can then
be exponentially close to the real axis, and the norm of the 
resolvent be of order $e^{C/h}$ for $z\in [-\delta,\delta]$ \cite{TZ,Bu02,CaVo04}.

In this note we are considering an intermediate situation, namely the case where the trapped set
\eqref{e:trapped} is a (locally maximal) {\it hyperbolic set}. 
This means that 
$K$ is a compact, flow-invariant set with no fixed point, 
such that at any point $\rho\in K$ the 
tangent space splits into the  neutral ($\RR H_p ( \rho) $), stable ($E_\rho^- $), and unstable
($E_\rho^+$) directions:
\[ T_\rho p^{-1} ( 0 ) = \RR H_p ( \rho) \oplus E_\rho^- \oplus E_\rho^+ \,.\]
This decomposition is preserved through the flow. 
The (un)stable directions are characterized by the following properties:
$$
\exists\, \lambda>0,\quad \norm{d \exp tH_p(\rho)v}\leq C\,e^{-\lambda |t|}\norm{v},\quad
\forall\, v\in E^{\mp}_\rho,\ \pm t>0.
$$
Such trapped sets are easy to construct. The simplest case consists in a 
single unstable periodic orbit, but we will rather consider the more general case where 
$K$ is a {\it fractal} set supporting
a chaotic flow; such a set contains countably many periodic orbits, which are dense on 
the set of nonwandering points $NW(K)\subset K$ \cite{KaHa}.

Our results will depend on the ``thickness'' of the trapped set, defined
formulated in terms of a certain dynamical object, the {\it topological pressure}. 
We refer to \cite[\S 3.3]{NZ3} and texts on dynamical systems \cite{KaHa,Walters} 
for the general definition of the pressure, recalling only a definition valid in the 
present case.
Let $ f \in C^0( K ) $. Then the {\em pressure} of $ f $ with respect
to the Hamiltonian flow on $K$ is given by 
\begin{equation}
\label{eq:press}   \cP ( f ) \defeq
\lim_{T \rightarrow \infty } \frac 1 T \log \sum_{ T_\gamma < T }
\exp  \int_0^{T_\gamma } ( \exp t H_p)^* f (\rho_\gamma)\,dt \,, 
\end{equation}
where the sum runs over all periodic orbits $\gamma$ of periods $ T_\gamma\leq T $, and $\rho_\gamma$
is a point on the orbit $\gamma$. The function $f$ we will be using is a multiple of the 
(infinitesimal) unstable Jacobian of the flow on $K$:
$$
\varphi_+ ( \rho ) \defeq 
\frac{ d } { dt }\det (d \exp t H_p \restriction E^+_\rho ) |_{t=0}\,,\qquad \rho\in K\,.
$$
We can now formulate our main result:

\medskip

%\begin{thm}\label{thm:resol}  
\noindent {\bf Theorem.} {\em Suppose that $ P ( h ) $ satisfies \eqref{eq:Pu} 
or the more general assumptions of \cite[\S 3.2]{NZ3}. Suppose
also that the Hamiltonian flow
is hyperbolic on the trapped set $ K $, and that the topological pressure
\begin{equation}
\label{eq:pre}   \cP ( - \varphi_+ /2 ) < 0 \,,\qquad \varphi_+\ \text{the unstable Jacobian}.
\end{equation}
Then for any $ \chi \in \CIc( \RR^n ) $ and $ \epsilon > 0 $,
%<|\cP ( - \varphi_+ /2 ) |$, 
there exist $\delta(\eps)>0$ and $h(\eps)>0$ 
such that the cut-off resolvent
$ \chi ( P ( h ) - z ) ^{-1} \chi$,
$ \Im z > 0 $, continues analytically to the strip
\[   \Omega_\epsilon ( h ) \defeq \left\{ z \; : \;  
{\Im z} >  h ( \cP ( - \varphi_+ /2 ) + \epsilon  ) \,, \ \ 
| \Re z | < \delta ( \epsilon ) \right\} \,,  \quad  0 < h < h ( \epsilon ) \,. 
\]
For $z\in\Omega_\epsilon ( h )\cap \{\Im z\leq 0\}$, this resolvent is 
polynomially bounded in $ h $:
\begin{gather}\label{eq:reses}
\begin{gathered}
\| \chi  (P( h )-z)^{-1} \chi \|_{ L^2\to L^2} \leq C(\eps,\chi)\, 
h^{-1 +c_{E}\,{\Im z}/{h}}\,\log (1/h)\,,\\
c_{E}\defeq \frac{n}{2|P ( - \varphi_+ /2 )+\eps/2|}\,.
\end{gathered}
\end{gather}
}
%\end{thm}

\medskip

%We used the continuity of the pressure with respect to the energy in order to simplify the statement
%-- see \cite[Theorem 3]{NZ3} for a slightly more precise formulation.
For any $s\in [0,1]$, the pressure $\cP(-s\varphi_+)$ measures relative strengths of
the complexity of the flow on $K$ (i.e. the number of periodic orbits), and the 
instability of the trajectories (through the Jacobian). For $s=0$, $\cP(0)$ only measures the complexity, it is 
the topological entropy of the flow, which is generally positive.
On the opposite, $\cP(-\varphi_+)$ is negative, it represents the ``classical decay rate'' of the flow.
The intermediate value $\cP(-\varphi_+/2)$ can take either sign, depending on the
``thickness'' of $K$.
In dimension $ n=2 $ the condition \eqref{eq:pre} is equivalent to the
statement that the Hausdorff dimension of $ K \subset p^{-1} ( 0 ) $
is less than $ 2 $. Since the energy surface $ p^{-1} ( 0 ) $ has
dimension $ 3 $ and the minimal dimension of a non-empty $ K $ is
$ 1 $, the condition means that we are less than ``half-way'' and $ K $
is {\em filamentary}. Trapped sets with dimensions greater than $ 2 $ are
referred to as {\em bulky}.

The first part of the theorem is the main result of \cite{NZ3},
see Theorem 3 there. Here we use the techniques developed in that paper
to prove \eqref{eq:reses}. 
For the Laplacian outside several convex
obstacles on $\IR^n$ (satisfying a condition guaranteeing strict hyperbolicity 
of the flow) with Dirichlet or Neumann boundary condition,
the theorem was proved by Ikawa \cite{Ik}, with the pressure being
only implicit in the statement.% which gave an explicit condition on distances and sizes of the obstacles. 
For more recent developments in that setting see \cite{NBu},\cite{Pest}, and \cite{NSZ}. 

In particular, for $z$ on the real axis the bound
\eqref{eq:reses} gives
\be
\label{eq:blog}  
\| \chi ( P ( h ) - z )^{-1} \chi  \|_{ L^2  \rightarrow 
L^2 } \leq 
C\, \frac { \log\left( 1 / h \right) } h \,,\quad z\in [-\delta(\eps),\delta(\eps)],\quad
0<h<h(\eps)\,.
\ee
This result was already given in \cite[Theorem 5]{NZ3} with a less
direct proof. It has been generalized to a larger class of manifolds
in \cite{KD} and \eqref{eq:reses} provides no new insight in that
setting.  

One of the applications of \eqref{eq:blog} in the case of the Laplacian 
is a {\it local smoothing} with a minimal
loss \cite{HC1} in the Schr\"odinger evolution (see \cite{Bu} for the original application in the
setting of obstacle scattering):
$$
\forall\, T>0,\forall\, \eps>0,\exists\, C=C(T,\eps),\qquad 
\int_0^T \norm{\chi e(-it\Delta_g)u}^2_{H^{1/2-\eps}}\;dt\leq C\,\norm{u}^2_{L^2}\,.
$$
One can also deduce from \eqref{eq:blog} a Strichartz estimate \cite{HC1,BGH}
%$$
%\norm{\exp(-it\Delta_g)u}_{L^p((0,T);L^q(\IR^n))}\leq C\,\norm{u}_{L^2(\IR^n)},\qquad 
%\frac{2}{p}+\frac{n}{q}\geq \frac{n}{2},\ \ p>2\,,
%$$
useful to prove existence of solutions for some related semilinear Schr\"odinger
equations.

In the case of the Laplacian ($V\equiv -1$),
the estimate in a strip \eqref{eq:reses} has important consequences regarding
the energy decay for the wave equation -- see \cite{Bu,HC2,GN09}
and references given there. In odd dimension $n\geq 3$, it implies that the
local energy of the waves decays exponentially in time.
The same type of energy decay (also involving a pressure condition)
has been recently obtained by Schenck in the setting of the damped wave equation on a 
compact manifold of negative curvature \cite{Sch}. 

\medskip

To prove \eqref{eq:reses} we use several methods and intermediate results from \cite{NZ3}.
Using estimates from \cite[\S 7]{NZ3}, we show in \S \ref{re} how to obtain a good parametrix for 
the complex-scaled operator, which leads to 
an estimate for the resolvent. As was pointed out to us by Burq
the construction of the
parametrix for the outgoing resolvent was the, somewhat implicit, key
step in the work of Ikawa \cite{Ik} on the resonance gap for
several convex obstacle. That insight lead us to re-examine the 
consequences of \cite{NZ3}.

We follow the notation of \cite{NZ3} with precise references given
as we go along. For the needed aspects of 
semiclassical microlocal analysis \cite[\S 3]{NZ3} and the
references to \cite{DiSj} and \cite{EZ} should be consulted.

\section{Review of the hyperbolic dispersion estimate}
\label{oops}

The central ``dynamical ingredient'' of the proof is a certain dispersion estimate 
relative to a modification of $P(h)$, which we will now describe.

The first modification of $ P ( h ) $ comes from the method
of complex scaling reviewed in \cite[\S 3.4]{NZ3}. For any fixed 
sufficiently large 
$ R_0 > 0 $, it results in 
the operator $ P_\theta ( h ) $, with the following properties.
To formulate them 
put 
\be
\label{eq:Ot} \Omega_\theta \defeq [-\delta , \delta ] + i[-\theta/C, C] \,, 
\ \ \theta = M_1 h \log ( 1/ h ) \,. \ee
Then 
\begin{eqnarray}
& P_\theta ( h ) - z \; : \; H_h^2 ( \RR^n ) \longrightarrow 
L^2 ( \RR^n) \ \ \text{is a Fredholm operator for $ z \in \Omega_\theta $,}
\label{eq:PFr} \\
& \forall \, \chi \in \CIc ( B ( 0 , R_0 ) ) \,, \ \ 
\chi R ( z, h )  \chi = \chi  R_\theta (z, h ) \chi \,. 
\label{eq:RPt} 
\end{eqnarray}
Here and below we set the following notation for the resolvents:
\[ R_\bullet ( z, h ) \defeq ( P_\bullet ( h ) - z ) ^{-1} \,, \ \ \Im z > 0 \,,  \]
and \eqref{eq:RPt} shows the meromorphic continuation of $ \chi R ( z , h) 
\chi $ to $ \Omega_\theta $, guaranteed by the 
Fredholm property of $ P_\theta ( h ) - z $.

The operator $ P_\theta ( h ) $ is further modified by an exponential 
weight, $ G^w =G^w ( x, h D) $, 
\[  G \in \CIc ( T^* \RR^n ) \,, \ \
 \supp G \subset p^{-1} ( ( - 2 \delta, 2 \delta ) ) \,, \ \ 
 \partial^\alpha G = {\mathcal O} ( h \log ( 1/ h ) ) \,, \]
where $ \delta > 0 $ is a fixed small number. The modified operator is obtained by conjugation:
\be\label{eq:pthep}
P_{\theta , \eps } ( h ) \defeq  e^{-\eps G^w / h } P_\theta ( h ) 
e^{ \eps G^w / h } \,, \ \ \eps = M_2 \theta\,, \ \
\theta = M_1 h \log ( 1/ h )  \,.
\ee
This operator has the same spectrum as $ P_\theta ( h ) $ 
and the following properties:
\begin{eqnarray}
& \nonumber
\text{if } \ \psi_0 \in S( T^*\IR^n ),\quad \supp \psi_0 \subset p^{-1} (  ( -3\delta/2, 3\delta/2 ) ) \,, \,,\\
& \label{eq:Pos}
\text{if }\quad \Im \psi_0^w ( x, h D) \, P_{\theta, \epsilon } ( h ) \, \psi_0^w ( x , h D) \leq C\, h \, \\
%& \label{eq:PRc}
%R_{\theta,\eps} (z, h )  =  e^{-\eps G^w / h } \, R_{\theta} (z, h )\, e^{\eps G^w / h } \,,
%, \\
%& \label{eq:estG} \| \exp( \pm \eps G^w / h ) \|_{ L^2  \rightarrow 
%L^2  } = {\mathcal O} ( h^{ - C M_2 } ) \,.
\end{eqnarray}
The main
reason for introducing the weight $G$ is to ensure the bound \eqref{eq:Pos}.
The specific choice of $ G $ is explained in \cite[\S 6.1]{NZ3}.
In particular $ G $
vanishes in some neibhbourhood of the trapped set $K$, and the operator
$\exp (\eps G^w ( x, h D ) )$ is an $h$-pseudodifferential operator $ B^w ( x , h D) $, with
symbol satisfying
\[ B\in h^{-N} S_\delta ( T^* \RR^n ) \,, \quad 
B\restriction_{\complement \supp G }  = 1  + {\mathcal O}_{S_\delta } ( h^\infty ) \,.
\]
As a result, if the spatial cutoff 
$\chi$ is supported away from $\pi\supp G$, calculus of semiclassical pseudodifferential
operators ensures that
\be\label{e:cutoff-resolv}
\chi R(z,h) \chi=\chi R_{\theta,\eps} (z, h )  \chi
+ \cO_{L^2\to L^2}(h^\infty)\norm{R_{\theta,\eps} (z, h )} \,.
\ee
From now on our objective will then be to estimate the norm $\norm{R_{\theta,\eps} (z, h )}_{L^2\to L^2}$.

We consider a final modification of $ P_{\theta,\eps}(h) $ near the zero
energy surface. Let $ \psi_0 \in S(T^*\IR^n) $ be supported in 
$ p^{-1} (  ( -3 \delta/2, 3 \delta /2 ) $ and equal to $ 1 $ in 
$ p^{-1} ( - \delta, \delta ) $. Define
\be
\label{eq:wtP} \widetilde P_{\theta, \epsilon } ( h ) \defeq 
\psi_0 ^w ( x, h D )\, P_{\theta, \epsilon }\, \psi_0^w ( x, h D ) \,, \ee
and the associated propagator
\be\label{eq:Ut} 
U ( t ) \defeq 
\exp \{ - it \widetilde P_{\theta, \epsilon } ( h )/h \} \,.
\ee
The crucial ingredients in proving \eqref{eq:reses} are 
good upper bounds for the norms
$$
\|U(t)\psi^w(x,hD)\|_{L^2\to L^2}\,,  \quad \text{on time scales}\quad  0\leq t\leq M\log(1/h) \,, 
$$
where 
$M>0$ is fixed but large, and
\be\label{e:psi}
\psi\in S(T^*\IR^n),\quad  \supp\psi\subset p^{-1}((-\delta/2,\delta/2)),\quad \psi=1\ \ \text{on}\ \ 
p^{-1}((-\delta/4,\delta/4))\,. 
\ee
From the bound
\eqref{eq:Pos} on the imaginary part of $\widetilde P_{\theta, \epsilon } ( h )$,
we obviously get an exponential control on the propagator:
\begin{equation}\label{e:uniform-bad}
\norm{U(t)}_{L^2\to L^2}\leq \exp(C\,t),\qquad t\geq 0\,.
\end{equation}
The reason to conjugate $P_\theta$ with the weight $G^w$ was indeed to ensure this exponential bound.
Together with the hyperbolic dispersion bound \eqref{eq:Hyp}, this exponential bound 
would suffice to get a polynomial bound $\cO(h^{-L})$ in \eqref{eq:reses}, for some (unknown) $L>0$.
To obtain the explicit value,
\[ -1+\frac{c_E \Im z}{h} \,, \]
for the exponent, we need to 
improve \eqref{e:uniform-bad} into the following uniform bound:
\begin{lem}\label{}
Let $\psi$ satisfy the conditions \eqref{e:psi}. 
Then, there exist $h_0,\,C_0>0$ such that, 
\be\label{e:uniform}
\| U ( t) \psi^w ( x, h D ) \|_{L^2\to L^2}\leq C_0\,,\quad 0\leq t\leq M\log(1/h),\ \ h<h_0\,.
\ee
\end{lem}
Before proving this Lemma, we state the major consequence of our dynamical assumptions
for the classical flow on $K$, namely its hyperbolicity and the ``filamentary'' nature of 
$K$ (expressed through \eqref{eq:pre}). It is a 
{\it hyperbolic dispersion estimate} which was explicitly written only in a 
model case \cite[Proposition 9.1]{NZ3} \cite{NZ3}, but 
can be easily drawn from \cite[Proposition 6.3]{NZ3}, 
in the spirit of \cite[\S 6.4]{NZ3}
As above, we take $\psi$ as in \eqref{e:psi}. For any $\eps>0$ we set 
$\lambda \defeq -\cP ( \varphi_+ / 2 )+\eps/2 $. For any $0<h<h(\eps)$, we then have
\begin{gather}
\label{eq:Hyp}
\begin{gathered}
\| U ( t ) \psi^w ( x, h D ) \|_{L^2\to L^2} \leq C\, h^{-n/2} \exp ( - \lambda t ) + 
\cO ( h^{M_3} ) \,, \\
\text{uniformly in the time range}\quad 0 < t < M \log ( 1/ h )\,.
\end{gathered}
\end{gather}
The constant $M$ is arbitrarily large, and $ M_3 $ can be taken as
large as we wish, provided we choose $ M_1 $ in \eqref{eq:Ot} 
large enough depending on $M$.
If the pressure $\cP ( \varphi_+ / 2 )$ is negative, one can take $\eps$ small enough
to ensure $\lambda>\eps/2>0$. The above estimate is then sharper than \eqref{e:uniform}
for times beyond the {\it Ehrenfest time} 
\begin{equation}\label{e:Ehren}
t_{E}\defeq c_{E}\,\log(1/h),\qquad c_{E}\defeq \frac{n}{2\lambda}\,.
\end{equation}
The large constant $M$ will always be chosen (much) larger than $c_{E}$.

\bigskip

\noindent {\it Proof of Lemma \ref{e:uniform}}. To motivate the proof 
we start with a heuristic 
argument for the bound \eqref{e:uniform}.
As mentioned above, the exponential bound \eqref{e:uniform-bad} is due 
to the fact that the imaginary part of $\widetilde P_{\theta, \epsilon } ( h )$
can take positive values
of order $\cO(h)$ (see \eqref{eq:Pos}). However, 
the construction of the weight $G$
shows that outside a bounded region of phase space of the form 
$$
V_{\rm pos}=p^{-1}(( - 2 \delta, 2 \delta ))\cap T^*_{\{R_1<|x|<R_2\}}\IR^n\,,
$$
the imaginary part of $\widetilde P_{\theta, \epsilon } ( h )$ is negative
up to $ {\mathcal O}(h^\infty ) $ errors.

The radius $R_1$ above is large enough, so that $V_{\rm pos}$ lies at {\it finite distance} from the trapped set.
As a result, any trajectory crossing the region $V_{\rm pos}$ will only spend a bounded time
in that region. For this reason, the propagator $U(t)$ on a large time $t\gg 1$ 
will ``accumulate'' exponential growth 
during a uniformly  bounded time only.

We now provide a rigorous proof, using ideas and results from \cite[\S 6.3]{NZ3}. 
The phase space $T^*\IR^n$ is split using a smooth partition of unity:
$$
1=\sum_{b=0,1,2,\infty} \pi_b \,,\qquad\pi_b\in C^\infty(T^*\IR^n,[0,1])\,.
$$
These four functions have specific localization
properties:

\begin{itemize}
\item $\supp \pi_b\subset p^{-1}((-\delta,\delta))$ for $b=0,1,2$

\item $\pi_\infty$ is localized outside $p^{-1}((-3\delta/4,3\delta/4))$

\item $\pi_{1}$ is supported near $K$, in particular, its support does
not intersect $V_{\rm pos}$

\item $\pi_2$ 
is supported away from $K$ but inside $\{|x|<R_2+1\}$

\item $\pi_0$ is supported near spatial infinity,
that is on $\{|x|>R_2-1\}$ where the operator $\widetilde P_{\theta, \epsilon } ( h )$ is {\it absorbing}
(the  imaginary part of its symbol is negative).

\end{itemize}

Employing a positive (Wick) quantization scheme (see for instance
\cite{Le}, and for the semiclassical setting 
\cite[\S 3.3]{KPS}), 
$\Pi_b={\rm Op}_h^+(\pi_b)$, we
produce a quantum partition of unity 
\[ \Id=\sum_{b=0,1,2,\infty}\Pi_b \,, \qquad \| \Pi_b \| \leq 1 \,.\]
The evolution $U(t)$ is then split between time intervals of length 
$t_0$, where $t_0>0$ is large but independent of $h$. 
Using the partition of unity, we decompose the propagator
at time $t=Nt_0$ into
$$
U(Nt_0)\,\psi^w(x,hD)=\big(\sum_{b=0,1,2,\infty} U_{b}\big)^N\,\psi^w(x,hD),
\quad \text{where}\quad U_{b}\defeq U(t_0)\,\Pi_{b}\,. 
$$
Expanding the power, we obtain a sum of terms $U_{b_N}\cdots U_{b_1}\psi^w$; to understand each such
term semiclassically, we investigate whether there exist true classical trajectories following
that ``symbolic history'', namely sitting in $\supp\pi_{b_1}$ at time $0$, in $\supp\pi_{b_2}$ at time $t_0$, etc.
up to time $Nt_0$.

Since the energy cutoffs 
$\psi$ and $\pi_\infty$ have disjoint support, no classical trajectory can spend time in both supports.
As a result, any sequence containing at least one index $b_i=\infty$ is {\it irrelevant} 
(meaning that the corresponding term is $\cO_{L^2\to L^2}(h^\infty)$) \cite[Lemma 6.5]{NZ3}.

Since any classical trajectory can travel in $\supp \pi_2$ at most for 
a finite time $\leq N_0 t_0$ before escaping,
\cite[Lemma 6.6]{NZ3} shows that the {\it relevant} sequences $b_1 \cdots b_N$ are of the form 
$$
b_i=1\quad\text{ for }\quad N_0<i<N-N_0\,.
$$ 
They correspond to trajectories spending most of the time near $K$. 
One then has
$$
U(Nt_0)\,\psi^w(x,hD)=U(N_0t_0) \,(U_1)^{N-2N_0}\,
U(N_0t_0)\,\psi^w(x,hD) + \cO_{L^2\to L^2}(h^{ M_5} )\,,
$$ 
uniformly for any $2N_0\leq N<M\log (1/h)$, where $M_5>0$ is large if the previous $M,\,M_i$ are.

Finally, using the fact that the weight $G$ vanishes on $\supp\pi_1$, \cite[Lemma 6.3]{NZ3} shows that
$$
U_1= U(t_0) \Pi_1 = U_0(t_0) \Pi_1+\cO_{L^2\to L^2}(h^\infty)\,,
$$
where $U_0(t_0)=\exp(-it_0 P(h)/h)$ is {\it unitary}. Hence, $\|U_1\|\leq 1+\cO(h^\infty)$,
while $\norm{U(N_0t_0)}$ is estimated using \eqref{e:uniform-bad}. 
$\hfill\square$

%We reiterate that \eqref{eq:Hyp} is the main estimate in \cite{NZ3}.
%It was inspired by the work of Anantharaman and the first author
%\cite{An},\cite{AnNo}, and crucial aspects of its proof are similar in 
%spirit to the arguments of Ikawa \cite{Ik}.

%%%%%%%%%%%%%%%%%%%%%%%%%%%%%%%%%
%%%%%%%%%%%%%%%%%%%%%%%%%%%%%%%%%
\section{Resolvent estimates}\label{re}
%%%%%%%%%%%%%%%%%%%%%%%%%%%%%%%%%
%%%%%%%%%%%%%%%%%%%%%%%%%%%%%%%%%

We can now prove the resolvent estimate \eqref{eq:reses} by 
constructing a parametrix for $ P_{\theta, \epsilon} ( h ) -z $,
$ z \in \Omega_\epsilon ( h ) $ defined in the statement of the
theorem. We will use the notation 
\[ \zeta \defeq z/h \]
 to shorten 
some of the formul{\ae}.
We want to find an approximate
solution to 
\[ (  P_{\theta, \epsilon } ( h ) - z ) u = f \,, \ \ f \in L^2 ( \RR^n ) \,,
\ \ z  \in \Omega_\epsilon ( h ) \,.  \]
First, the ellipticity away from the energy surface $p^{-1}(0)$ shows that, for $\psi$ as in 
\eqref{e:psi}, there
exists an operator, $ T_0 = {\mathcal O} ( 1 ) : L^2 ( \RR^n ) 
\rightarrow H_h^2 ( \RR^n ) $, such that                          
\[ (  P_{\theta, \epsilon } ( h ) - z )  T_0 f = ( 1 - 
\psi^w ( x, h D) ) f + R_0 f\,,\qquad R_0 = \cO_{L^2\to L^2} ( h^\infty )\,.\]    
To treat the vicinity of $p^{-1}(0)$ we put
\[ T_1 f = (i/h) \int_0^{t_M} dt\,e^{i\zeta t}\, U ( t)\, \psi^w ( x, h D ) f \,, \quad t_M=M\log (1/h)\,,
\] 
which satisfies          
\be
\label{eq:Ptt}
 ( \widetilde P_{\theta, \epsilon } ( h ) - z )\, T_1 f =
 \psi^w ( x, h D) f +R_1 f\,,\qquad R_1\defeq  - e^{i\zeta t_M}\,U(t_M)\,\psi^w ( x, h D) \,.
\ee
The estimate \eqref{eq:Hyp} shows
that, if $\lambda + \Im\zeta>\eps/2$, and for arbitrary $M_4>0$, 
one can choose $M$ and $M_3$
large enough such that $R_1 = \cO_{L^2 \to L^2 } ( h^{M_4} )$.
We can estimate the norm of $T_1$ by the triangle inequality,
\be\label{e:triang}
\norm{T_1}_{L^2\to L^2}\leq h^{-1}\int_0^{t_M}e^{-\Im \zeta t}\, 
\norm{U(t)\,\psi^w ( x, h D)}_{L^2\to L^2} \, dt \,,
\ee
and then use the bounds \eqref{e:uniform} for times $0\leq t\leq t_{E}$
and \eqref{eq:Hyp} for times $t_{E}< t\leq t_{M}$.

When $\Im\zeta=0$, the above
integral can be estimated by the integral over the interval $t\in [0, t_{E}]$: 
$$
\Im\zeta=0\Longrightarrow \norm{T_1}_{L^2\to L^2}\leq h^{-1}\Big(C_0\,t_{E} + \frac{1}{\lambda}\Big)
\leq C\, h^{-1}\,\log h^{-1}\,.
$$
In the case $0>\Im\zeta>-\lambda+\eps/2$, the dominant part of 
the integral
comes from $t=t_{E}$:
$$
0>\Im\zeta>-\lambda+\eps/2 \Longrightarrow \norm{T_1}_{L^2\to L^2}\leq C_{\eps} h^{-1}\,e^{-\Im\zeta t_{E}}=
C_{\eps}\,h^{-1+c_{E}\Im\zeta}\,.
$$
We rewrite \eqref{eq:Ptt} as
\[  \psi_0^w ( x, h D )  (  P_{\theta, \epsilon } ( h ) - z )  
\psi_0^w ( x, h  D) T_1 f = \psi^w ( x, h D ) f + R_1 f \,. \]
From the inclusion $ \psi_0 \rest_{\supp \psi} \equiv 1 $, one can show
(as in \cite[Lemma 6.5]{NZ3}) that
$$
\psi_0^w ( x, h D )  (  P_{\theta, \epsilon } ( h ) - z ) \psi_0^w ( x, h  D) T_1= 
(  P_{\theta, \epsilon } ( h ) - z )\,T_1+R_2\,,\qquad R_2 = \cO_{L^2\to L^2 } ( h^\infty )\,,
$$
and also that
$$
\norm{T_1}_{H^2_h}\leq C\,\norm{T_1}_{L^2}\,.
$$
Putting $ T = T_0 + T_1 $ and $ R = R_0+ R_1 + R_2 $, we obtain
\[ (  P_{\theta, \epsilon } ( h ) - z ) T = \Id + R \,, \ \ 
\quad  R =  \cO_{L^2 \to L^2  } ( h^{M_4} ) \,.\] 
This means that $ (  P_{\theta, \epsilon } ( h ) - z ) $ 
can be inverted, with
$$
\norm{(  P_{\theta, \epsilon } ( h ) - z )^{-1} }_{L^2\to H_h^2}= (1+\cO(h^{M_4})) 
\norm{T}_{L^2\to H_h^2}\,.
$$ 
The above estimates on the norms of $T_0$ and $T_1$ can be summarized by
\be
0\geq\Im\zeta\geq \eps+\cP(-\varphi_+/2)\Longrightarrow \norm{T}_{L^2\to H_h^2}\leq 
C_{\eps}\,h^{-1+c_{E}\Im\zeta}\,\log h^{-1}\,.
\ee
Using \eqref{e:cutoff-resolv}, this proves the bound \eqref{eq:reses}.
$\hfill\square$

\medskip

\noindent {\bf Remark.} By using a sharper energy cutoff $\psi_h$ belonging to 
an exotic symbol class (see \cite[\S 4]{SjZwA}) and 
supported in the energy layer $p^{-1}((-h^{1-\delta},h^{1-\delta}))$
(as in \cite{AnNo}), the bound \eqref{eq:Hyp} is likely to be improved to 
\begin{equation}\label{e:improved}
\| U ( t ) \psi_h^w ( x, h D ) \|_{L^2\to L^2} \leq C h^{-(n-1+\delta)/2} \exp ( -  \lambda t ) + {\mathcal O} ( h^{M_3} )\,.
\end{equation}
This bound becomes sharper than \eqref{e:uniform} around the time $t'_{E}=c'_{E}\log (1/h)$, where
$$
c'_{E}\defeq\frac{n-1+\delta}{2\lambda}<c_{E}.
$$
As a result, the bounds on the norm of the corresponding operator $T'_1$ 
are modified accordingly.
At the same time, as shown in  \cite[Prop.~5.4]{AnNo}, the ellipticity 
away from the energy surface provides an 
operator $T'_0$ satisfying 
$$
(  P_{\theta, \epsilon } ( h ) - z )  T'_0  = ( 1 - \psi_h^w ( x, h D) ) +\cO_{L^2\to L^2}(h^\infty)\,,
$$
and of norm $\norm{T'_0}_{L^2\to L^2} =\cO(h^{-1+\delta})$. The norm of $T'=T'_0+T'_1$
is still dominated by that of $T'_1$, so that we eventually get
$$
\norm{\chi (P(h)-z)^{-1}\chi }_{L^2\to H_h^2}\leq C_{\eps}\,h^{-1+c'_{E}\Im z/h}\,\log (1/h)\,,
\quad z\in \Omega_\epsilon ( h )\cap \{\Im z\leq 0\}.
$$
Since it is not clear that even this bound is optimal, and that proving 
\eqref{e:improved} would require some effort, we have limited
ourselves to using the established bound \eqref{eq:Hyp}. 

One advantage of
the approach presented in this note (compared with the method of \cite[\S 9]{NZ3})
is that, to obtain the
bound \eqref{eq:blog} we did not have to use the complex interpolation
arguments of \cite{Bu} and \cite{TZ}.

\section{Funding}
This work was supported by
the Agence Nationale de la Recherche [ANR-05-JCJC-0107-01, S.N]; and
the National Science Foundation [DMS 0654436, M.Z.].

\acks{In addition to Nicolas Burq, we would like to thank
Nalini Anantharaman and Jared Wunsch for helpful discussions related to 
\cite{NZ3}.}

\end{document}